\documentclass{PoS}

\PoS{PoS(LAT2005)008}

\title{Hadronic Decays}

\ShortTitle{Hadronic Decays}

\author{\speaker{C. Michael}\\
Theoretical Physics Division, Dept of Mathematical Sciences,  
   University of Liverpool, Liverpool L69 3BX, UK\\

\email{c.michael@liv.ac.uk}
 }

\abstract{ Hadronic decays and transitions are a key ingredient  of
hadronic physics. I discuss how hadronic decays can be explored in lattice
gauge theory  and review studies undertaken. I also discuss the impact
of decays on masses  and how lattice studies can explore the nature of
a hadronic state: namely whether it is a molecular or quark-antiquark state.
A brief discussion of lattice exploration of pentaquark states is presented.
 }

\FullConference{XXIII International Symposium on Lattice Field Theory\\  
                25th-30th July 2005\\
	        School of Mathematics, Trinity College, Dublin}

\usepackage{color}
\let\normalcolor\relax

\newcommand{\black}{\color{black}}
\newcommand{\blue}{\color{blue}}
\newcommand{\red}{\color{red}}
\newcommand{\magenta}{\color{magenta}}

\usepackage{graphicx}

\begin{document}

\section{Introduction}

  Relatively few hadronic states are stable to strong decays (i.e. via
QCD with degenerate $u$ and $d$ quarks).  Among the mesons, we
have~\cite{Eidelman:2004wy}:

\bigskip
\begin{tabular}{ll}
 {\black Stable} &{ \blue $\pi\ K\ \eta\ D\ D_s\ B\ B_s\ B_c\  D_s^*\ B^*\ B_s^*\
  D_s(0^+)\  B_s(0^+)$}   \\ 
 {\black $\Gamma < 1$ MeV} &{ \red
$\eta'\ D^*\ \psi(1S)\  \psi(2S)\ \chi_1\ \chi_2\
\Upsilon(1S)\ \Upsilon(2S)\ \Upsilon(3S)$}\\ 
{\black $\Gamma < 10$ MeV }&{ \magenta 
$ \omega\ \phi\ \chi_0\ X(3872) $} \\
 {\black $\Gamma > 10$ MeV }&{ \red  $ \rho\ f_0\ a_0\ h_1\ b_1\ a_1\
f_2\ f_1\ a_2,$\ \black etc., inc \red $\eta_c$.}    
 \end{tabular}

\bigskip

The mass of an unstable state is usually defined as the energy
corresponding  to a 90$^0$ phase shift. This definition seems to accord
with simple mass formulae:
  For example
 \begin{itemize}
 \item  $\rho(776)$ and  $\omega(783)$ are close in mass despite having 
widths of $150$ and  $8 $ MeV respectively.
 \item The baryon decuplet ($\Delta(1232)$,  $\Sigma(1385)$,  $\Xi(1530)$,
 $\Omega(1672)$) is roughly equally  spaced in mass despite having widths
of (120, 37, 9, 0) MeV respectively.
 \end{itemize}
 (Note that defining the mass as the real part of the pole will cause a
downward shift  of masses for wider states, eg. 22 MeV
less~\cite{Michael:1966aa} for the $\Delta(1232)$ pole, and this
prescription  will fit the equal mass rule less well.)

\bigskip

 So, on the one hand, unstable particles seem to  fit in well with
stable ones;  on the other hand, the presence of open decay channels
will have an influence in  lattice studies.

\bigskip

Some of the motivations to study hadronic decays on the lattice are:

\begin{itemize}
  \item To determine properties of exotic states (glueball; hybrid meson;
multi-quark) to guide experiment.
  \item To understand the nature of states: whether meson-meson or
quark-antiquark in structure.
  \item To understand the impact of decay on the mass of a state.
  \item Hadronic decays are now accessible in full QCD: 
 for example $\rho_0 \to \pi_1 \pi_{-1}$ at rest requires 
 $$ {m(\pi) \over m(\rho) } < {0.5 \over \sqrt{1+{4 \pi^2 \over
(m(\pi)L)^2}}} $$
 so for $m(\pi) L = 5$ this implies $m(\pi)/m(\rho) < 0.32$,
 but note that $\rho_1 \to \pi_1 \pi_0$ , where suffix labels momentum
in units of $2\pi/L$, only needs $m(\pi)/m(\rho) = 0.44$,

\end{itemize}

\section {Decays in Euclidean Time} 

  {\magenta NO GO.} At large spatial volume, the two-body continuum
{\red masks}  any resonance state.
        The extraction of the spectral function from the correlator $C(t)$ is 
ill-posed unless a model is made~\cite{Michael:1989mf,Maiani:1990ca},
 since the low energy continuum dominates at large $t$.

\bigskip

     {\magenta GO}.   For finite spatial volume ($L^3$), the two-body
continuum is  {\red discrete} and L\"uscher
showed~\cite{Luscher:1986a,Luscher:1986b,Luscher:1990ux} how to use the
small energy  shifts with $L$ of these two-body levels to extract the
elastic scattering phase shifts. The phase shifts then determine the
resonance mass  and width, see ref.~\cite{Luscher:1991cf} for a review. 
    { Thus a relatively broad resonance such as the $\rho$
appears as a distortion of the $\pi_n \pi_{-n}$  energy levels
where pion momentum $q=2 \pi n/L $.}   

 The effect can be visualised as arising from the relatively larger 
amplitude for interaction between two hadrons in increasingly smaller
spatial volumes:

 \vspace{3cm}

\includegraphics{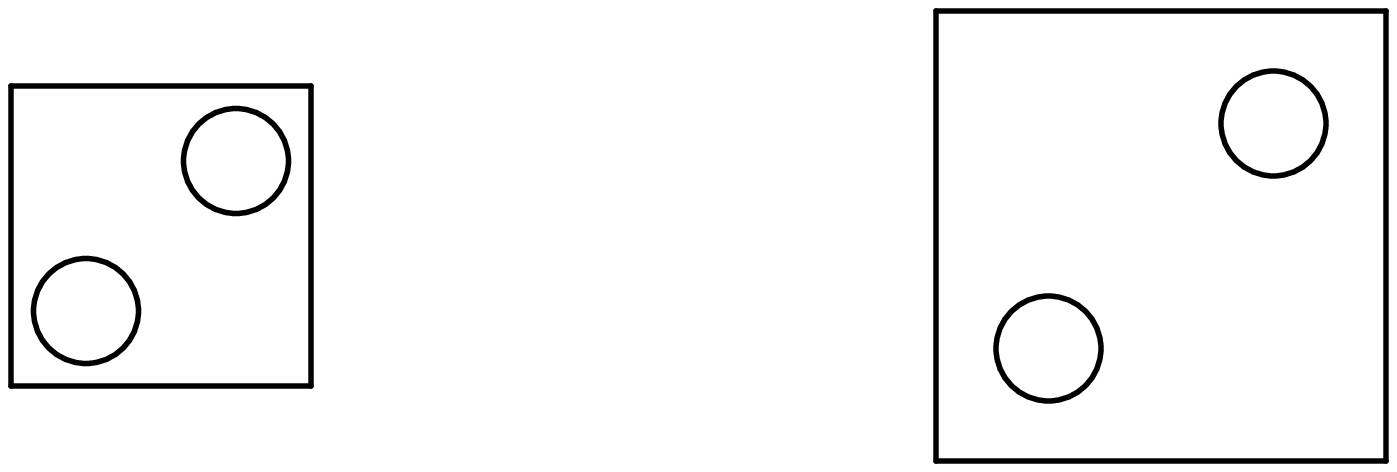}

\section{Lattice evaluation}

{ L\"uscher's method is applicable in principle to study phase shifts below 
inelastic thresholds.  But problems arise
in practical implementation:
 \begin{itemize}
 \item Accurate measurement of a small energy difference (of order
$L^{-3}$ in general, though bigger at resonance) is needed.
 \item Measurement of a matrix of  correlations between two-body and
one-body operators  will be needed to get accurate energy values and
these correlators are less straightforward to evaluate.
 \item Measurements are needed at several spatial sizes ($L$)  with
dynamical fermions, with $L$ big enough to hold two hadrons.
 \item Higher energy levels in a given channel are particularly hard to
determine (eg. $\pi_1 \pi_{-1}$ compared to $\pi_0 \pi_0$) since excited
 state energies are always difficult to extract with precision from 
temporal correlators, since the ground state contribution dominates at
larger $t$.
 
\end{itemize}

\noindent  Are there any short cuts?

\section {Lattice evaluation: tricks}

\subsection{$\pi^+ \pi^+$}

One of the first cases that has been explored is the $\pi^+ \pi^+$
interaction at low energy. This is favourable since there  are fewer
quark diagrams to evaluate as there is no annihilation diagram, i.e no
$\overline{q} q $ channel and, because of this,  quenched evaluation is a
useful  approximation. This has been much studied in the last decades. 
Indeed recently there have also been  dynamical quark
evaluations~\cite{Yamazaki:2004qb,Beane:2005rj} of the  $\pi^+ \pi^+$
scattering near threshold.

The trick that has been evolved to improve precision in these studies is
that of  using an operator that creates two pions at spatial separation
$x$. Then the nature of the expected spatial wave function versus $x$ 
allows to determine the scattering length (phase shift $\delta_{I=2}$
near threshold) more accurately than from the energy determined by the
$t$-correlation alone~\cite{Aoki:2005uf}.

This $\pi^+ \pi^+$ case does not involve decay, of course, just an
evaluation of the  hadronic interaction strength.  To explore decays,
one must study the transition between two meson and one meson operators.

\subsection{Hadronic transitions}

Consider a lattice study of the off-diagonal correlator: from a $\rho$ meson to 
$\pi \pi$. Diagrammatically:

\ \ \ \ \  {\red $\rho$} $\to$ {\blue $\pi \pi$} \\
\ \ \ 0{\red ----------}X{\blue----------}0 \\
\ \ \ 0 \ \ \ \ \ \ \ \ \ \ \ t\ \ \ \ \ \ \ \ \ \ \ \ \ T

\includegraphics{fd_tri.ps}

Now to evaluate this contribution, since the intermediate point marked X
at time $t$ is not observed on a lattice, it must be summed over.
   $$ \sum_{t=0}^T {\red e^{-m(\rho)t}} x {\blue e^{-m(\pi \pi)(T-t)}}
 \to   A e^{-m(\rho)T} - B e^{-m(\pi \pi)T}$$

The problem that arises is that excited states of either 
the $\rho$ or of $\pi \pi$ will contribute a similar  behaviour:

    $$ \sum_{t=0}^T {\red e^{-m'(\rho)t}} x {\blue e^{-m(\pi \pi)(T-t)}}
 \to   C e^{-m'(\rho)T} - D e^{-m(\pi \pi)T}$$

    $$ \sum_{t=0}^T {\red e^{-m(\rho)t}} x {\blue e^{-m'(\pi \pi)(T-t)}}
   \to   E e^{-m(\rho)T} - F e^{-m'(\pi \pi)T}$$

Hence there is  no way in principle  to remove any  excited state
contamination {\magenta unless}
 ${\red m(\rho)} \approx {\blue m(\pi \pi)}$ when the ground state piece
 sums to $xTe^{-mT}$  while excited states only behave as $e^{-mT}$. 

 Thus for on-shell transitions on the lattice, it is possible to extract
the hadronic  transition amplitude
directly~\cite{McNeile:2000xx,Pennanen:2000yk,McNeile:2002fh}. The key
signal is to observe a linear dependence of the lattice normalised
transition  amplitude on the temporal extent $T$. The slope of this 
transition then gives the lattice amplitude $x$ that can be related to 
the transition amplitude with conventional normalisation of states.

\section {Lattice evaluation: $\rho \to \pi \pi$ }

  This transition can be evaluated directly - for {\red on
shell}   transitions -  by looking for the signal  {\red extensive} in $T$.
 This has been explored~\cite{McNeile:2002fh} for $\rho$ decay to  $\pi
\pi$. In order to have an on-shell transition with the dynamical  quark
lattices then available, it was optimum to study  the case of decay in
flight:  $\rho_1 \to \pi_1 \pi_0$ which is quite  close to on-shell as
illustrated in fig.~1. The diagrams illustrated in fig.~1 were evaluated
using a stochastic time-plane source method. The generalisation of
L\"uscher's  method to decay in flight is given
in~\cite{Rummukainen:1995vs}.

\begin{figure}[htb]
 \begin{center}
 \vspace{7.0cm} 

\includegraphics{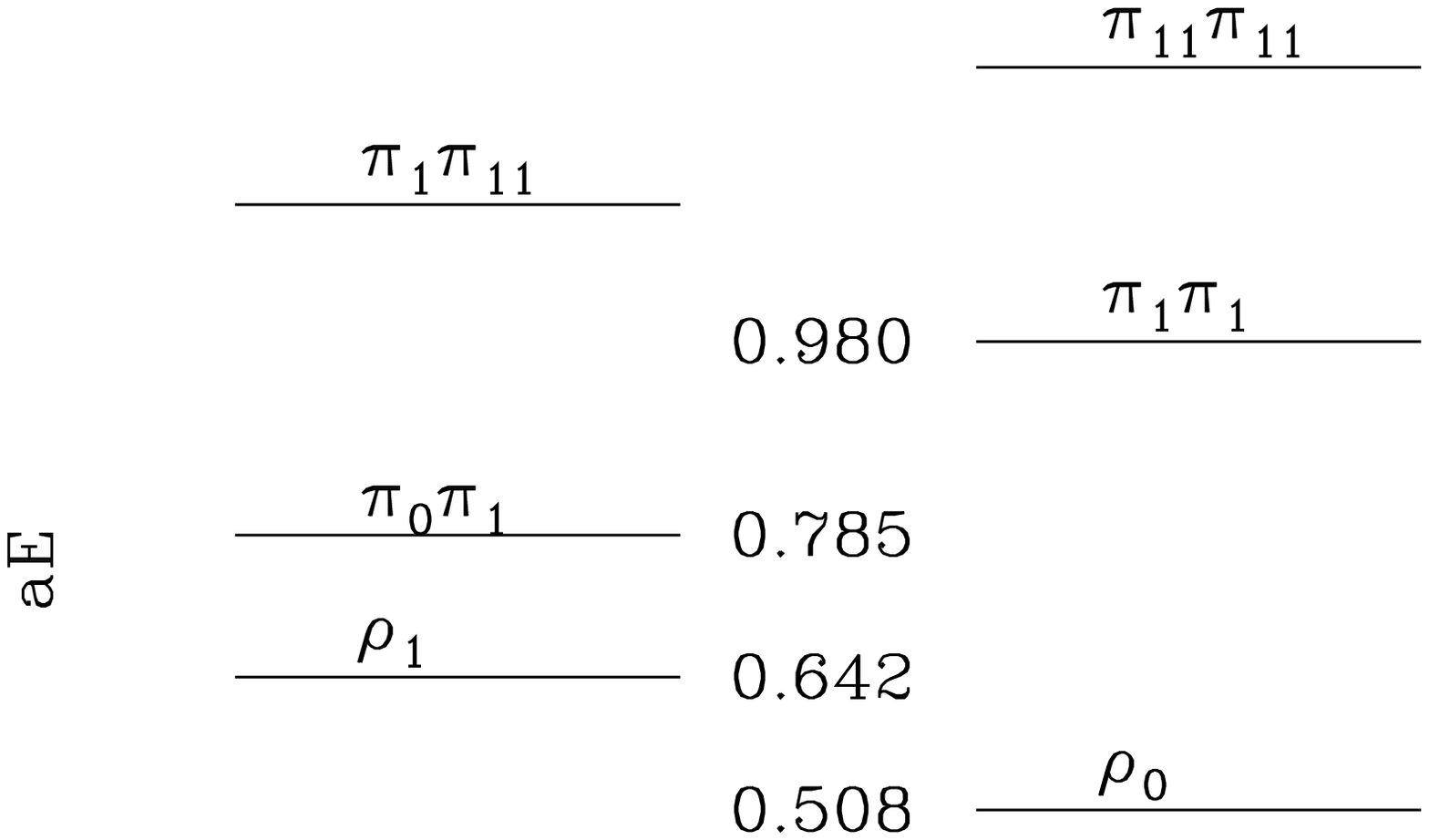}
\includegraphics{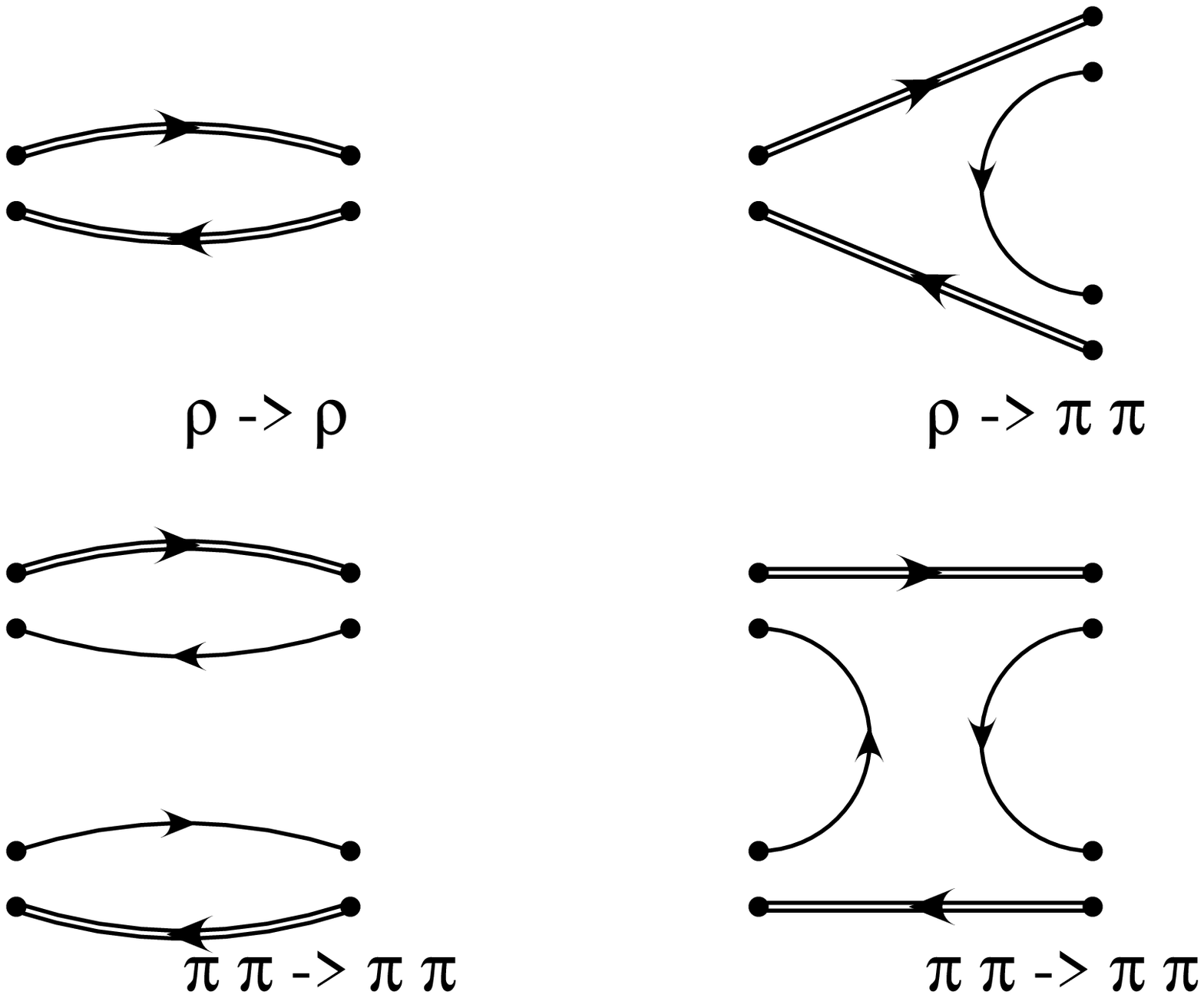}

 \end{center}
 \vspace{-2cm}
 \caption{(a) Energy levels of $\rho$ meson and $\pi \pi$ states.
 From UKQCD with $N_f=2$ sea quarks with NP-improved fermions, 
$m(\pi)/m(\rho)= 0.58$, $m(\pi)L=4.6$, $a$=0.11 fm. 
   (b) Diagrams evaluated (stochastic time-plane source method)
 }

 \end{figure}

The signal obtained from the lattice study is illustrated in fig.~2 both
for the  normalised $\rho \to \pi \pi$ transition and for the normalised
"box"  contribution $\pi \pi \to \pi \pi$ which has a contribution from
$\pi \pi \to \rho \to \pi \pi$. The approximate consistency of these two
approaches  is a useful cross-check.

\begin{figure}[hbt]

 \begin{center}
 \vspace{7.0cm} 

\includegraphics{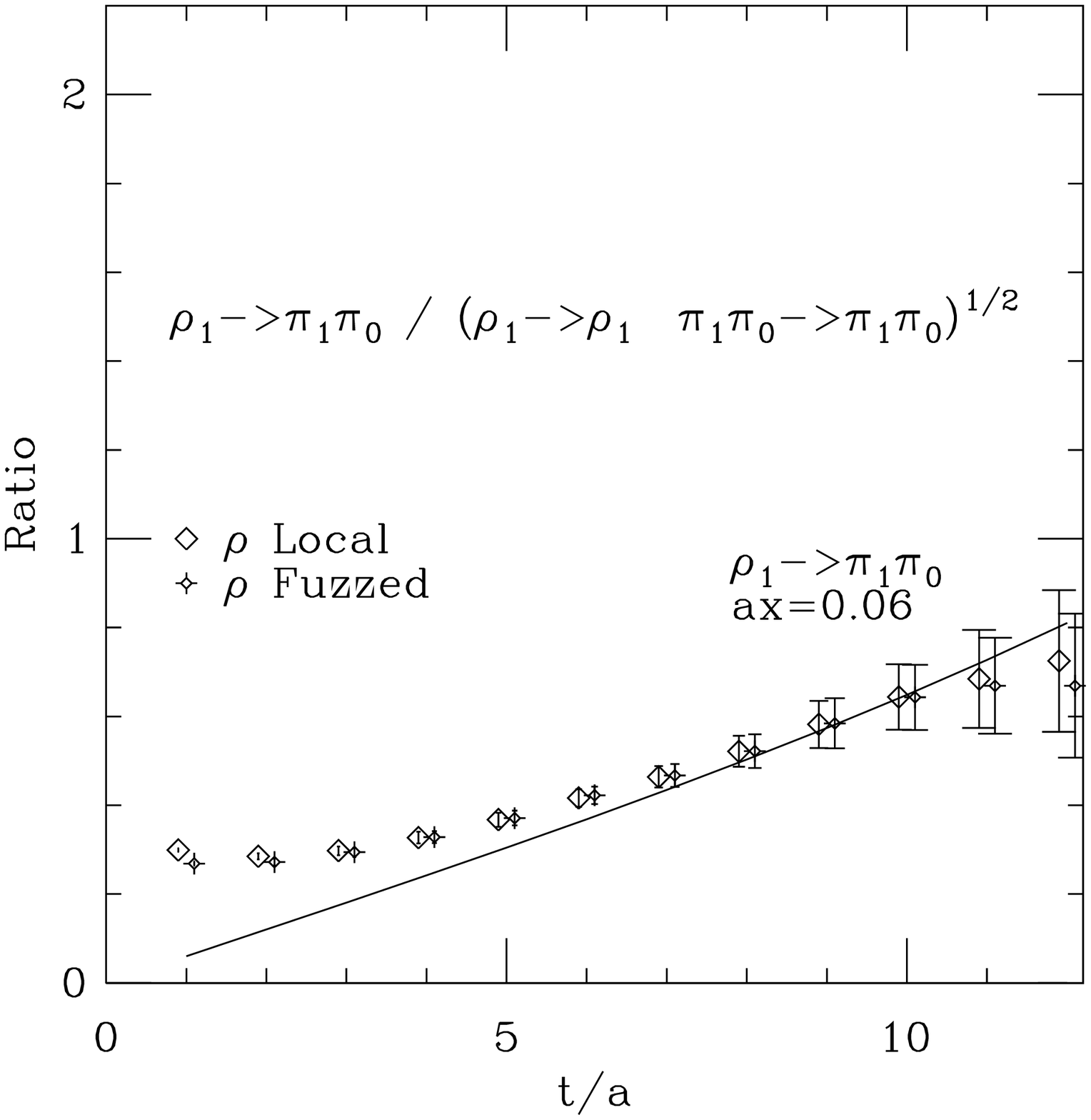}
\includegraphics{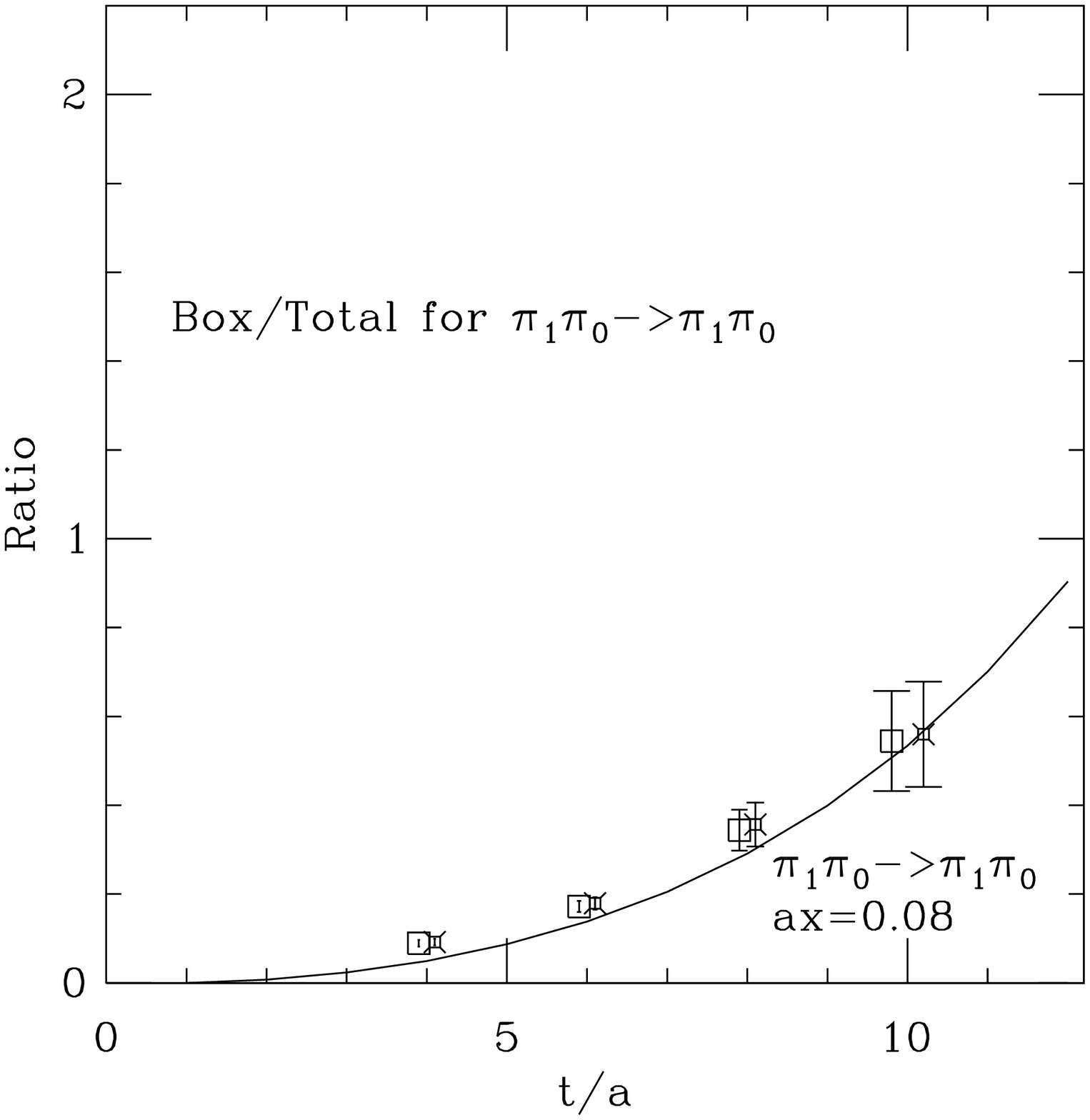}

 \end{center}
 \caption{Strength of transitions measured on the lattice.
 }

 \end{figure}

    Assuming two close energy levels (with mixing amplitude $x$) {\red
mix} to give  an energy shift $\Delta m$, then  $m(\rho_1)$ moves down,
while  $E(\pi_1 \pi_0)$ moves up. The lattice results are
 \begin{itemize}
   \item Transition $\rho_1 \to \pi_1 \pi_0$:  Signal $xT$. 
  {\red $a\Delta m=0.022^{+17}_{-7}$}
   \item Box $\pi_1 \pi_0 \to \pi_1 \pi_0$:  Signal $ (xT)^2$.
(consistent with above)
   \item Shift of energy of $\rho_1$ spin parallel to momentum (which
{\red mixes} with $\pi_1 \pi_0$) to spin perpendicular to momentum
(which does not).
   {\red $a\Delta m =0.026(7)$}
   \item L\"uscher shift (unbinding energy of $\pi_1 \pi_0$). 
 {\red $a\Delta m =0.04(3)$}. {\magenta Note big error}
    \end{itemize}

 So the tricks give smaller errors than the direct determination of the
L\"uscher mass shift. The tricks have some inherent systematic errors, 
however, and are most useful for a qualitative study.

  These mass shifts with the assumption of $\rho$ dominance of the $\pi
\pi$ partial wave give a determination of the coupling $g_{\rho \pi \pi}$.
 Thus one can determine this coupling constant from the lattice (where decay
does not proceed)  and compare with experiment:

 \bigskip
\begin{tabular}  {llll}
 \hline
  method  &     $m_{val}$  & $m_{sea}$ & $\overline{g}$ \\ 
 \hline
 {\red Lattice} $xT$ &         $s$   & $s$    &  $1.40^{+47}_{-23}$ \\ 
 {\red Lattice} $\rho$ shift &         $s$   & $s$    &  $1.56^{+21}_{-13}$ \\ 
 \hline
 $\phi \to K \overline{K}$ &         $s$   & $u,\ d$    &  $1.5$ \\ 
 $K^* \to K \pi$ &         $u,\ d/s$   & $u,\ d$    &  $1.44$ \\ 
 $\rho \to \pi \pi$ &         $u,\ d$   & $u,\ d$    &  $1.39$ \\ 
 \hline
\end{tabular}
 \bigskip

 The agreement is good, indicating that there is a relatively mild
dependence of the  coupling constant on the sea quark mass (which is
higher in the lattice study than in experiment). This satisfactory
confrontation  between lattice and experiment, is encouraging for
lattice exploration of cases where predictions need to be made since
experimental data  are not available.

\section {Hybrid meson decay}

 One of the characteristic predictions of QCD is that there can be
mesons in which the gluonic degrees of freedom  are non-trivially
excited. The simplest example is  a hybrid meson with  spin-exotic
$J^{PC}=1^{-+}$ which is a $J^{PC}$ combination not available to  a
$\overline{q} q$ state. 

\subsection{Heavy quarks}

\begin{figure}[htb]
 \begin{center}
 \vspace{7.0cm} 

\includegraphics{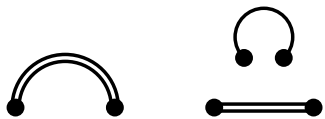} 
\includegraphics{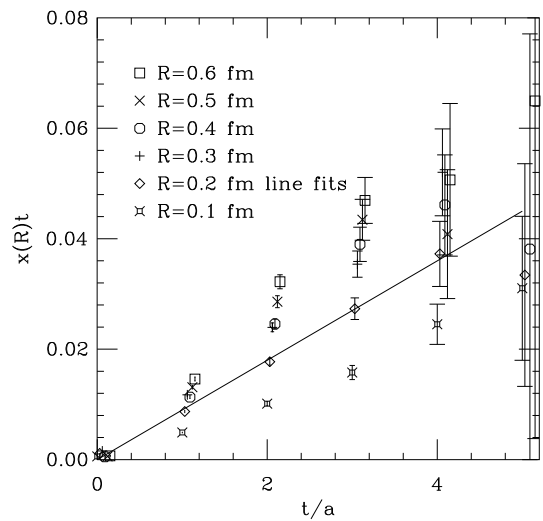}

 \end{center}
 \caption{Strength of string de-excitation transitions measured on the lattice.
 }

 \end{figure}

The cleanest environment in which to study such states on a lattice is 
in the limit of very heavy quarks - $\overline{b} b$. This can be
approximated by using  static quarks and the gluonic excitation arises
as an excited string state  between these static quarks with non-trivial
gluonic angular momentum. Lattice studies have long predicted the
spectrum of such states. 

To guide experiment, however, it is important to know the expected decay
 mechanism and associated width. In the static quark limit, several 
symmetries can be used which imply~\cite{McNeile:2002az} that the
dominant  decay will be string de-excitation (rather than string
breaking). Lattice study~\cite{McNeile:2002az} shows that the  dominant
decay of $H_b$  is string de-excitation  to  $\chi_b f_0$.  The
transition that is considered is shown by the diagram on  the left of
fig.~3. The lattice data on the transition are illustrated in fig.~3: 
the transition is closest to on-shell for
 $R \approx 0.2$ fm. The  width is predicted to be around 80 MeV.  

 This  estimate from first principles of the decay width is of 
significance in guiding experimental searches for such hybrid states.


\subsection {Light quarks}

  For light quarks, evidence for contributions from two-body states
(such as $\pi b_1$) to the  spin-exotic ($J^{PC}=1^{-+}$) channel is seen
 in dynamical studies~\cite{Bernard:2003jd} - this complicates the
extraction of a spin-exotic hybrid meson.
  

 A recent exploratory study (quenched) of $\pi a_1$ mixing with the
$J^{PC}=1^{-+}$  spin-exotic hybrid meson has given an estimate of the
width for this decay~\cite{Cook:2005aa}.

\section {String breaking}

 In quenched QCD, as the static quark and antiquark sources are  pulled
apart the potential energy rises as $\sigma R$, where $\sigma$ is the
string  tension. In full QCD, a sea quark-antiquark can be created from
the vacuum and it becomes  energetically favourable to have a
meson-antimeson pair beyond some value of $R$. This is the phenomenon of
string breaking.  It has long been realised that this can be studied as
a {\red mixing} phenomenon~\cite{Michael:1991nc}  with channels    $Q
\overline{Q}$ and  $ Q\overline{q}\   q\overline{Q}$. 
   For static quarks at separation $R$, there will be a level crossing 
and associated mixing of $V(R)$ and $2m(B)$. This mixing is the measure 
of string breaking~\cite{Pennanen:2000yk},  \cite{Bernard:2001tz}. This
mixing energy is very hard to determine on a lattice:  since the
crossing occurs at relatively large $R \approx 1.25$ fm where amplitudes
are small. At this crossing point, for static quarks, there is  a mixing
transition  which will be independent of lattice spatial size $L$ for
sufficiently large $L$ since each state is localised (the heavy quarks
are static).

Employing "all the tricks in the book" a first estimate of this  energy
shift (mixing amplitude) has been obtained~\cite{Bali:2005fu}   of 
51(3) MeV. An illustration of the crossing region is in fig.~4.

Using the adiabatic approximation,  this energy gap  can be used to
evaluate amplitudes  for  excited $\Upsilon$ decay to $B\overline{B}$. So it
does indeed have some relevance to the topic of  hadronic decays.


\begin{figure}[htb]

 \begin{center}
 \vspace{7.5cm} 

\includegraphics{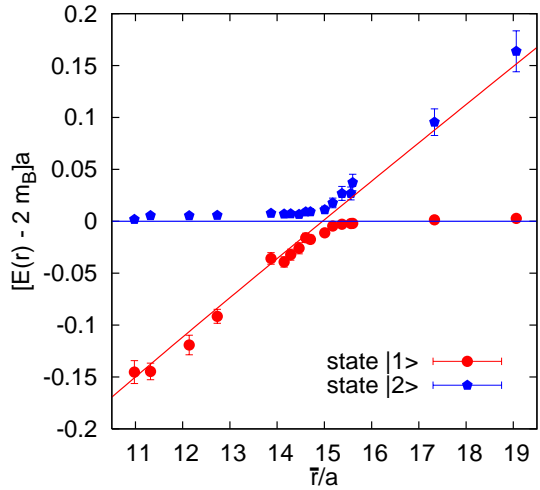} 

 \end{center}
\vspace{-1cm}
 \caption{String breaking (here $a \approx 0.083$ fm).
 }

 \end{figure}

\section {Scalar Mesons}

\subsection{Light quarks}

 Since $u\overline{u}+d\overline{d}$, $s\overline{s}$,  glueball, and meson-meson
components are all possible for flavour-singlet scalar mesons, this is 
a difficult area to study both on a lattice, and in interpreting
experimental data.  For scalar mesons the lowest  mass decay channels are
$\pi \pi$ (flavour singlet: $f_0$)  or $\eta \pi$ (flavour non-singlet:
$a_0$) and these decay channels are open in many dynamical lattice
studies. The history of attempts to study the complex mixing between
these  different contributions is :

 \begin{itemize}
   \item  $0^{++}$ Glueball decay $\to \pi \pi$: quenched
study~\cite{Sexton:1995kd,Sexton:1996ed}.
   \item Glueball mixing with $q \overline{q}$ meson. 
 This is a  hadronic transition which is independent of $L$ for large $L$, so 
a mixing energy can be quoted. 
  It has been studied using quenched~\cite{Lee:1999kv} and dynamical
lattices~\cite{McNeile:2000xx} 
  \end{itemize}

      A full lattice  study is needed which includes glueball, $\overline{q} q$ and $\pi \pi$ 
channels but the disconnected diagram for $f_0 \to \pi \pi$ 
is very noisy in practice. 
 To reduce the contribution from disconnected diagrams, one can study
flavour non-singlet decays. 
 The simplest case is $a_0 \to \eta \pi$ and this has been explored in
quenched studies  which have an anomalous behaviour: since the $\eta$
itself is unphysical (appearing as a double pole degenerate in mass
with the pion). Rather than try to correct for this anomaly which gives 
a wrong sign to the $a_0$ correlator at larger $t$, it is preferable to use 
a ghost-free theory. 
 With two flavours of sea quark ($N_f=2$), the $a_0 \to \eta \pi$
transition  is physical but it involves the evaluation of an additional
disconnected diagram -  as illustrated in fig.~5. This transition is
then approximately  on-shell and the transition strength can be
evaluated, as shown in fig.~5.

 The  results~\cite{cmcmjp} are indeed as anticipated for the transition
involving  particles with no momentum - showing an approximately linear
rise. For the  decay in flight the result is significantly different
even though, for an  S-wave transition, one would expect a transition
amplitude independent  of momentum.  This is perhaps a  warning that the
underlying dynamics is more complicated.  Further study is needed in this 
area of light-light scalar mesons.

\begin{figure}[htb]

 \begin{center}
 \vspace{6.5cm} 

\includegraphics{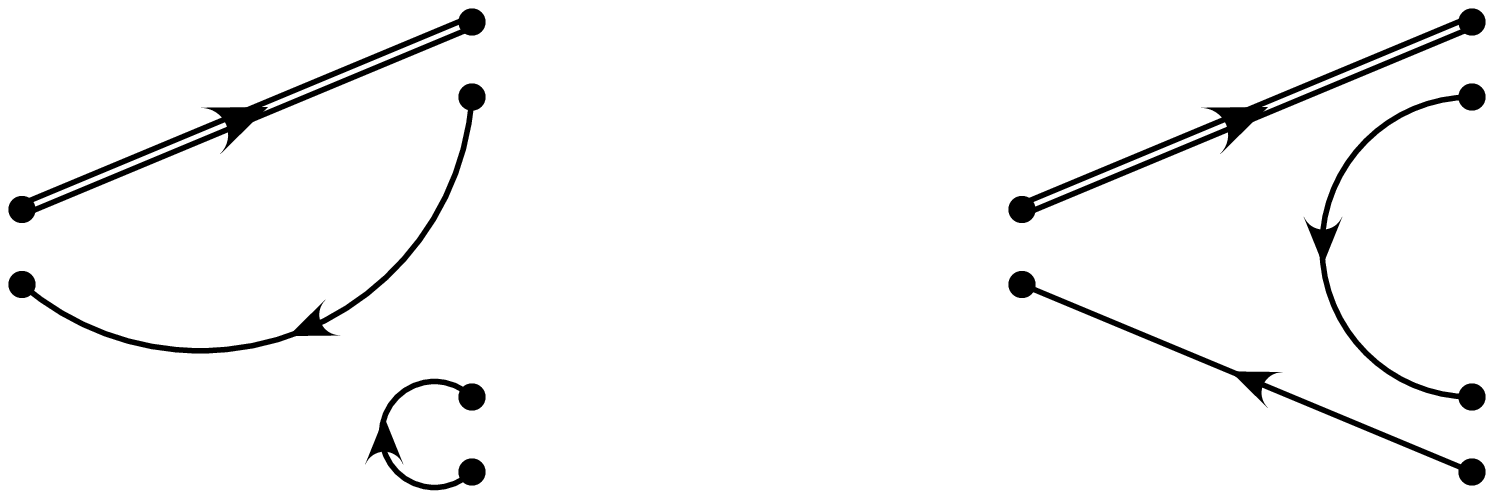} 
\includegraphics{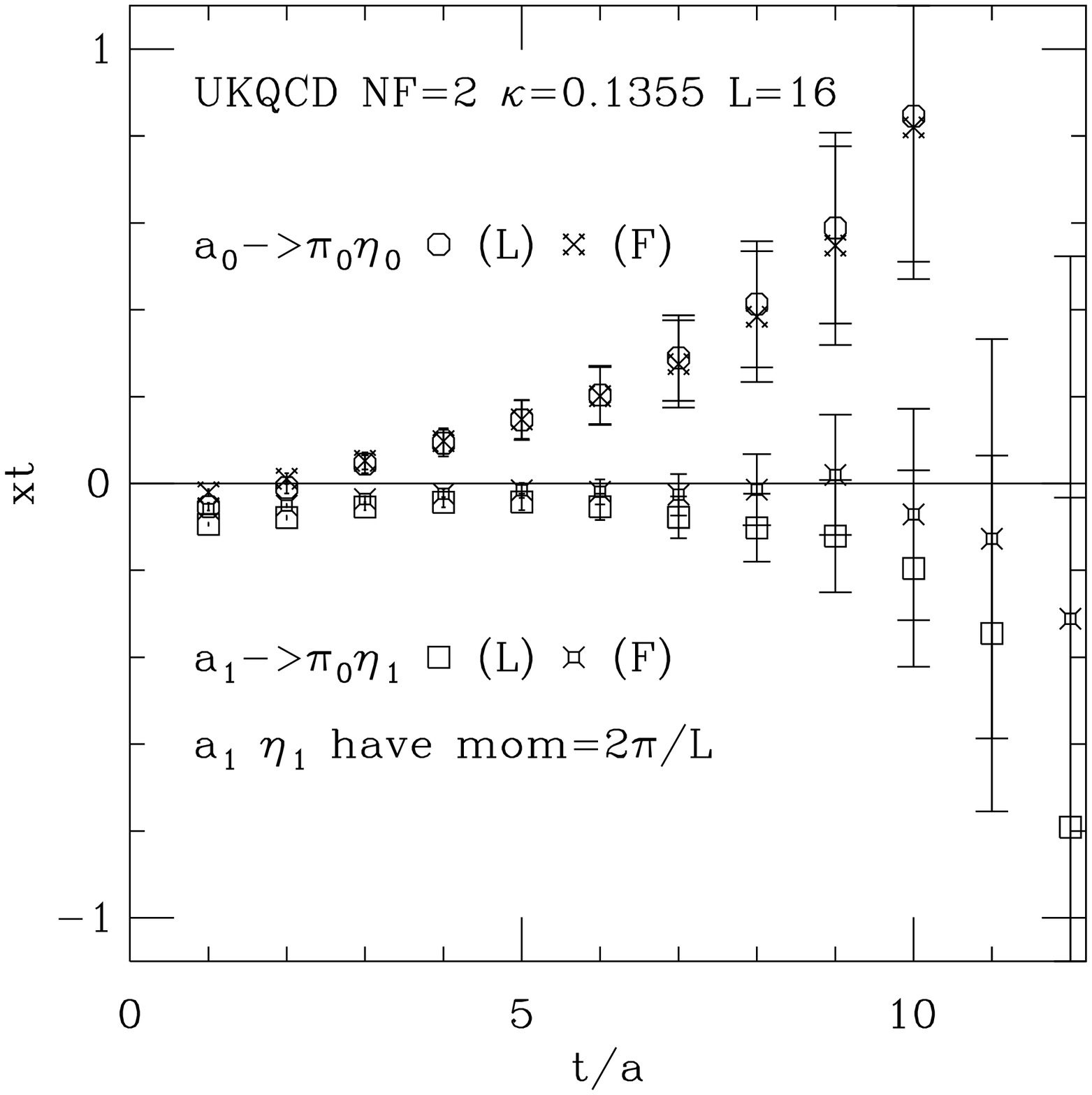} 

 \vspace{-0.5cm}
 \end{center}
 \caption{ Diagrams for $a_0 \to \eta \pi$ and transition strength on a
lattice (preliminary results from McNeile et al.~\cite{cmcmjp}).
 }

 \end{figure}

 \subsection{Heavy-light quarks}

 One of the most promising areas to study scalar mesons on  lattice is
for  heavy-light mesons. The scalar meson with $\overline{c}s$ quantum
number is known  experimentally~\cite{Eidelman:2004wy} to be very narrow
(it decays only via the isospin-violating  channel $D_s \pi$ or
electromagnetically), while the scalar meson with  $\overline{b}s$
quantum number is predicted to be similarly narrow from a lattice study 
of its energy~\cite{Green:2003zz}. 

 The heavy-light scalar meson, which in the limit of a static heavy
quark, is  expected to be stable~\cite{Green:2003zz} for $\overline{b} s$
content and to decay to $B \pi$ for  $\overline{b}n$ content (where $n=u,\
d$, considered as degenerate).  
 Evaluating the diagram shown in fig.~6, a lattice estimate of the decay
rate  of  $B(0^+) \to B(0^-) \pi$   gives a width
predicted~\cite{McNeile:2004rf}  as  162(30) MeV. This state has not
been  observed experimentally yet, but the  experimental results for the
corresponding $\overline{c}n$ state, $D(0^+)$,  are that the width is 
$270\pm50$ MeV. Although significant  $1/m_Q$ effects are expected in
the HQET in extrapolating to charm quarks, this is indeed a similar
magnitude to that predicted for $B$ mesons. 
 It will be interesting so see how the lattice prediction of the  mass
and width of the $B(0^+)$ fares when experimental results are available.


\begin{figure}[htb]

 \begin{center}
 \vspace{7.5cm} 

\includegraphics{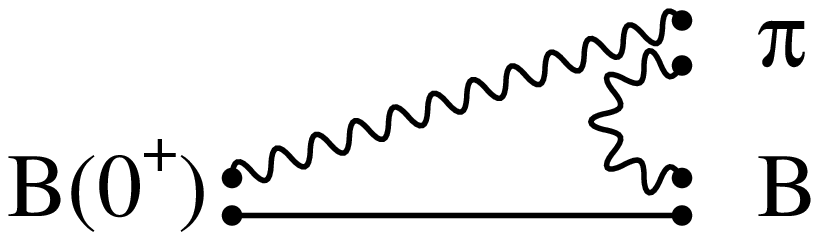} 
\includegraphics{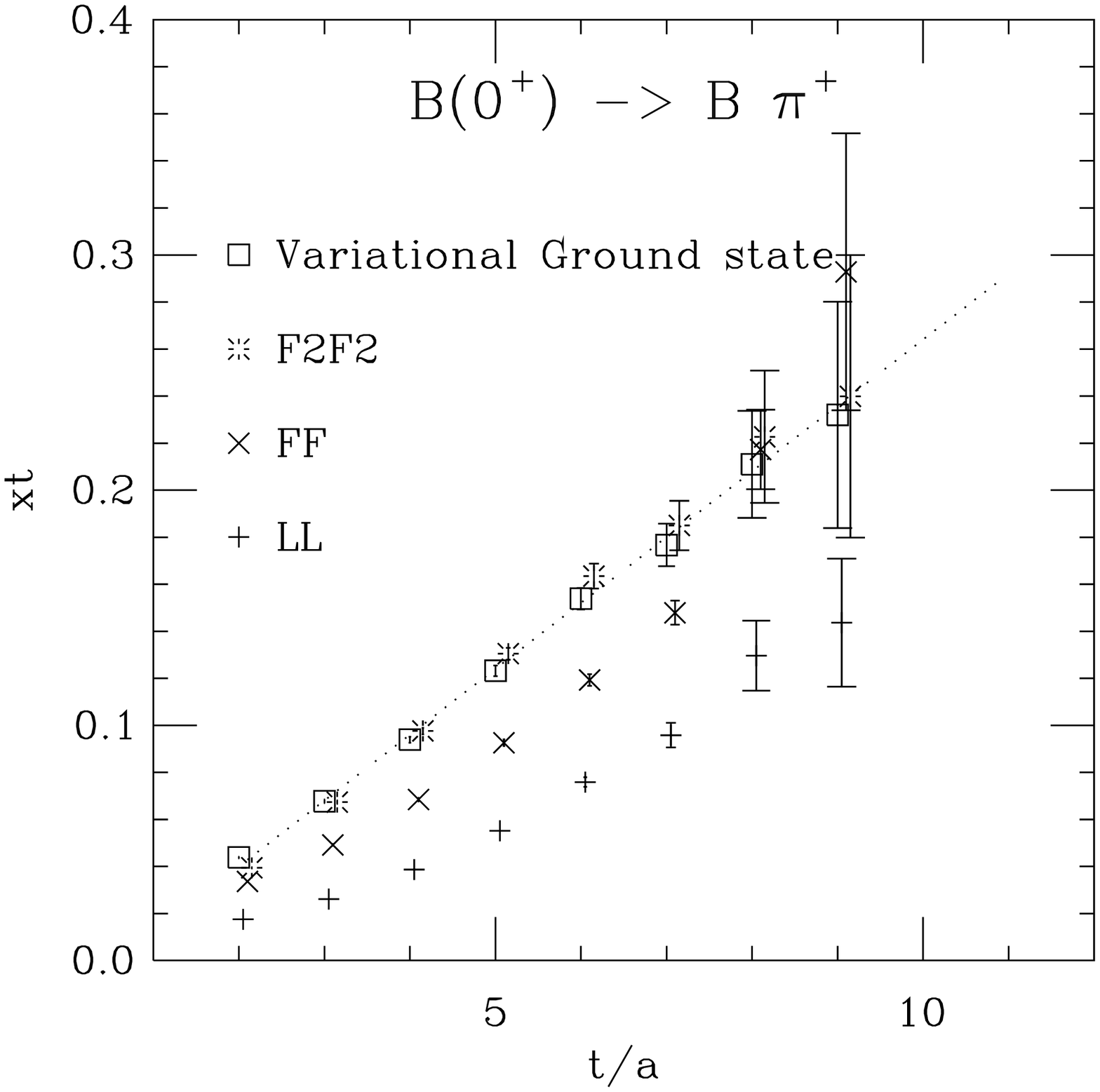}

 \end{center}
 \vspace{-1.0cm}  
 \caption{ Diagrams for $B(0^+) \to B  \pi$ and transition strength on a
lattice.
 }

 \end{figure}

\section {Do decays matter?}

  The previous discussion of decays on a lattice emphasises  that $q
\overline{q}$ states do mix with two-body states  with the same quantum
numbers. In the real world, the two-body states are a continuum  and
nearby states have a predominant influence.  Since there is a
suppression in the amplitude near threshold from  the factor $q^L$ for
an $L$-wave transition,  for S-wave transitions ($L=0$)  the threshold
will turn on most abruptly and hence will have stronger mixing.

For {\red bound} states there is an influence of nearby many-body states
(eg. $N \pi$ on $N$ or $\pi \pi \pi$ on $\pi$) which mix to reduce the
mass. The nearest such  thresholds will be those with pionic channels
since pions are the lightest mesons.  This is the province of low energy
effective theories, especially Chiral Perturbation Theory which is
discussed in other  talks. This then provides a reliable guide in
extrapolating lattice results to the  physical light quark masses.

For {\red unstable} states (resonances) the influence of the two-body
continuum  is less clear since the two-body states are both lighter and
heavier.  In the continuum at large volume, effective field theories 
can again be used to explore this. On a lattice, however, the signal for
 a particle becomes obscured as the quark mass is reduced so that it
becomes unstable. Techniques, such as those discussed above, are needed to 
extract the elastic scattering phase shift and hence the mass and width.

For quenched QCD, however, where these two-body states  are not coupled
(or have the wrong sign as in $a_0 \to \eta \pi$), then the  unstable
states will be distorted compared to full QCD. For instance, in existing
quenched QCD studies,  the  $\rho$ will be {\em too heavy} since it is
not repelled by the  heavier $\pi \pi$ states. Indeed an example of this
effect was seen above in the study of $\rho$ decay including dynamical
sea quarks, where the  $\rho$ mass decreased~\cite{McNeile:2002fh} when
it could couple to  $\pi \pi$ compared to  when it could not.

\section {Molecular states?}

 Can lattice QCD provide evidence about possible molecular states: 
hadrons made predominantly of two hadrons?

 The prototype is the deuteron: $n\  p$ bound in a relative S-wave (with
some D-wave admixture) by $\pi$ exchange.

 There are many states close to two-body thresholds. Since S-wave
thresholds are the  most abrupt, it is usually in this case that  the
influence of the  threshold on the state has been discussed. 
Some of these   cases are:

 \begin{tabular}{rclrcl}
&&&&& \\
$f_0(980)\ a_0(980)$ &$\leftrightarrow$& $K \overline{K}$ &
       $ D_s(0^+)$ &$\leftrightarrow$& $D(0^-) K$\\
       $ B_s(0^+)$ &$\leftrightarrow$& $B(0^-) K$ &
       $ X(3872)$ &$\leftrightarrow$& $D^* \overline{D}$\\ 
       $ \Lambda(1405)$ &$\leftrightarrow$& $\overline{K} N$ &
       $ N(1535)$ &$\leftrightarrow$& $\eta N$\\
&&&&& \\
 \end{tabular} 

Some of these states ($ D_s(0^+)$, $B_s(0^+)$) are stable (in QCD in the
isospin  conserving limit) whereas the rest have other channels open.  
There is a large literature, stretching over 40 years, discussing the 
consequences of the nearby threshold on these states. One definite 
implication is that  isospin breaking is enhanced by mass splittings in
thresholds (eg. $\overline{K^0}  K^0$ compared to $K^+ K^-$ is 8 MeV higher 
and this induces isospin mixing between the states at 980 MeV). 
This level of detail is not accessible in lattice studies at present, but 
lattice QCD should be able to address the issue of the influence of 
thresholds on these states.


 The observation of a state near a  2-body threshold implies that there 
is an attractive interaction between the two bodies.  But this is a
topic like that of whether the chicken or egg was created first:  an
attractive interaction implies and is implied by a nearby state. What
can lattice QCD offer here? We are in the position of being able to 
vary the quark masses and this is a very useful tool. A two-body
threshold will move in general in a different way with changing quark
mass than a $\overline{q}q$ state. We can also move the strange and
non-strange masses  separately and this can be helpful too.

Another line of investigation is that lattice studies can explore the 
wavefunction of a state - either the Bethe-Salpeter wavefunction or the 
charge or matter spatial distribution.  One can also explore the
coupling of a state to a 2-body channel, as was discussed  above. 

The prototype of a molecular state is the deuteron: it has a tiny
binding energy (2.2 MeV)  and a very extended spatial wave function.
Pion exchange between neutron and proton  gives a mechanism for this
long-range attraction. In general it is difficult to  reproduce such
small binding energies in lattice studies. 

Another case where a long-range pion exchange can give binding is in the
$BB$  system. Here lattice results indicate~\cite{Michael:1999nq} the
possibility  of molecular bound   states in some quantum number channels
which have  an attractive interaction from pion exchange,  but also the 
possibility of bound multi-quark states which are not described as
hadron-hadron  but where the two heavy quarks form a colour triplet and
the light quarks  are arranged as in a heavy-light-light baryon. This
$BB$ example illustrates the rich structure available to multi-quark 
systems. 

Cases that have been studied on the lattice are the $\overline{b}s$ and
$\overline{c}s$  scalar mesons~\cite{Green:2003zz,Dougall:2003hv}.  The
lattice mass values  do suggest that these states, treated as $\overline{Q}q$
states, are bound and  are not unduly influenced by the $BK,\ DK$
threshold. A study of the charge  distribution of the $B(0^+)$  
gives~\cite{Green:2004nm,Koponen:2005aa} confirmation  since the light
quark spatial distribution is similar to that of other  $\overline{Q}q$
states.   

For the $a_0$ and $f_0$ at 980 MeV, there is limited progress. Quenched 
studies are inappropriate here for the $a_0$ since the $\eta \pi$ decay
is  wrongly treated. In dynamical
studies~\cite{McNeile:2000xx,Bernard:2001av}, the two body channels are
of similar  energy to the $a_0$ (as discussed above), so analysis is
unclear. A thorough  study of the light-light scalar sector is still
awaited.

\section {Multi-quark states?}

{\red I told you so}: a narrow pentaquark above $KN$ threshold is not
possible in QCD. I lived through the {\red split $A_2$} and {\red
baryonium},  which were both narrow features that appeared significant
experimentally but which  were evanescent~\cite{baryonium}. So I  have
always advised lattice theorists not to rely on the experimental 
evidence for a pentaquark. A pentaquark state around 1540 MeV has a
"fall-apart"  decay mode to $KN$. This decay does not involve any quark
pair production so  is expected to be unsuppressed - resulting in a very
wide decay width (of the order  of hundreds of MeV). 

Since the flimsy experimental evidence for the pentaquark is now
weakening~\cite{Eidelman:2004wy,jlab,Abe:2005gy}, it is  less compelling
to study it exhaustively on the lattice. Nevertheless,   it is
instructive to discuss what lattice study could do to make more firm my 
conviction that "a narrow pentaquark is not possible in QCD".  The
essential issue is the narrowness of the claimed signal, since a broad
resonance state would not  be unexpected.
 In a  lattice exploration of the pentaquark,   an attractive phase
shift in some KN channel is not  sufficient to resolve the issue,  the  
{\em width} needs to be evaluated. This needs, in principle,  
\begin{itemize}
 \item 2+1 flavours of sea quark, with light $u$,$d$.
 \item Operators to create multi-quark states and two-body states
 \item Vary spatial size to determine phase shift of two-body interaction.
\end{itemize}

 Such a comprehensive study is not yet available. Instead most studies
have  used quenched QCD. The $KN$ decay channel {\em is} coupled in
quenched studies  to a pentaquark state, so this suggests that a
quenched study can be a useful  first step, although care must be  taken
of spurious contributions arising from states such as $KN\eta$ which
will be  relatively light and have unphysical behaviour.  For realistic
quark masses, there will  be two body  states on the lattice: their
energy shifts then, following L\"uscher, give the  phase shift. By
exploring this phase shift versus energy, the width of any  resonance
can then be extracted.   In practice, for a very narrow resonance, the 
spectrum will look more like a one-body state (the resonance) and a 
collection of two-body states except for mixing near the avoided-level
crossings. As the lattice volume is varied, the two body levels with
non-zero momentum will move (since momentum is  $2 \pi n/L$ ). Thus the 
volume dependence is a useful diagnostic. It is difficult, however,  to
determine accurately the many energy levels expected in a relatively
large volume. An additional indicator is that the weight of the 
one-body and two-body contributions can be volume dependent: basically 
because the contribution from a  two body state  to a local-local
correlator will be dominated by the lowest {\em relative} momentum which
has a contribution which  behaves as $1/L^3$ compared to the
contribution from  a one-body state.

Using these criteria, lattice groups have explored the S- and P-wave
$KN$ system. A summary of lattice results at LAT04~\cite{Sasaki:2004vz}
showed  most groups reproducing a pentaquark state.  Many more  results
have been presented since and at this conference. Here I comment  on
some of the more comprehensive studies.   The Kentucky
group~\cite{Mathur:2004jr} use relatively light quarks and  conclude no
evidence for any pentaquark. One group which claimed lattice evidence
for a pentaquark signal~\cite{Csikor:2003ng} has subsequently withdrawn
their claim~\cite{Csikor:2005xb}. The analysis of quenched lattices  at
quark masses heavier than physical is still somewhat subjective and
lattice evidence for a pentaquark is still being claimed by some groups,
e.g.~\cite{Alexandrou:2005gc}.

\section {Conclusions}

 Hadronic physics involves {\red unstable} states.

 Don't put your head in the sand: study these with lattice techniques.

 We need to study interactions as well as spectra: solid gold - not just
gold-plated.

\providecommand{\href}[2]{#2}\begingroup\raggedright\endgroup

\end{document}